\documentclass{article}
\usepackage{dcase2020,amsmath,graphicx,url,times,booktabs,tabularx}
\graphicspath{ {./images/} }
\usepackage{hyperref}
\usepackage{amssymb,bm}
\usepackage{siunitx}
\title{Temporal Sub-sampling of Audio Feature Sequences for Automated Audio Captioning}
\name{Khoa Nguyen$^{1}$,
      Konstantinos Drossos$^{2}$\thanks{K. Drossos and T. Virtanen would like to acknowledge CSC-IT Center for Science, Finland, for computational resources. Part of the computations leading to these results were performed on a TITAN-X GPU donated by NVIDIA to K. Drossos.},
      Tuomas Virtanen$^{2}$}
\address{$^1$ 3D Media Group, Tampere University, Tampere, Finland\\
        \{firstname.lastname\}@tuni.fi\\
        $^2$ Audio Research Group, Tampere University, Tampere, Finland\\
        \{firstname.lastname\}@tuni.fi}
\begin{document}
\begin{sloppy}
\ninept
\maketitle
\begin{abstract}
Audio captioning is the task of automatically creating a textual description for the contents of a general audio signal. Typical audio captioning methods rely on deep neural networks (DNNs), where the target of the DNN is to map the input audio sequence to an output sequence of words, i.e. the caption. Though, the length of the textual description is considerably less than the length of the audio signal, for example 10 words versus some thousands of audio feature vectors. This clearly indicates that an output word corresponds to multiple input feature vectors. In this work we present an approach that focuses on explicitly taking advantage of this difference of lengths between sequences, by applying a temporal sub-sampling to the audio input sequence. We employ a sequence-to-sequence method, which uses a fixed-length vector as an output from the encoder, and we apply temporal sub-sampling between the RNNs of the encoder. We evaluate the benefit of our approach by employing the freely available dataset Clotho and we evaluate the impact of different factors of temporal sub-sampling. Our results show an improvement to all considered metrics.
\end{abstract}
\section{Introduction}\label{sec:intro}
Audio captioning is the task of automatically describing the contents of a general audio signal, using natural language~\cite{drossos2017automated,lipping:2019:dcase}. It can be considered an inter-modal translation task, where the contents of the audio signal are translated to text~\cite{lipping:2019:dcase,wu:2019:icassp}. Audio captioning offers the ability for developing methods that can learn complex information from audio data, like spatiotemporal relationships, differentiation between foreground and background, and higher and abstract knowledge (e.g. counting)~\cite{lipping:2019:dcase,drossos2020clotho}.

Audio captioning started in 2017~\cite{drossos2017automated} and all published audio captioning methods (up to now) are based on deep neural networks (DNNs)~\cite{wu:2019:icassp,kim:2019:acl}. The usual set-up of methods is according to sequence-to-sequence architectures, where an encoder (usually based on recurrent neural networks, RNNs) takes a sequence of audio feature vectors as an input, and a decoder (also usually RNN-based) takes as an input the output of the encoder and outputs a sequence of words. A common element to sequence-to-sequence architectures is the alignment of the input and output sequences~\cite{cho2014learning,bahdanau2014neural,seo:2016:iclr}. Three main alternative techniques have been proposed for the alignment of the input and output sequences, namely the employment of a fixed-length vector representation of the output of the encoder~\cite{cho2014learning}, the usage of attention mechanism~\cite{bahdanau2014neural,seo:2016:iclr}, and self-attention~\cite{vaswani:2017:neurips}. The former two approaches have been widely adopted in the audio captioning field. Specifically,~\cite{drossos2017automated} presents a method that employs an RNN encoder, an RNN decoder, and an attention mechanism between the encoder and encoder, to align the input and output sequences. Study~\cite{wu:2019:icassp} used again an RNN encoder and an RNN decoder, but the alignment of the input and output sequences performed with the usage of a fixed-length vector. This vector was the mean, over time, of the output of the encoder. Finally,~\cite{kim:2019:acl} presented an approach for audio captioning, employing the attention mechanism presented in~\cite{seo:2016:iclr}. 

Although the above-mentioned techniques seem to be essential for the audio captioning task, they are applied only at the output of the encoder. This means that the encoder processes the whole input audio sequence, and only at its output there is the association (through the alignment mechanisms) of the different parts of the input sequence with the different parts of the output sequence. If we could adopt a policy for effectively collate sequential parts of the input sequence, then we might be able to let the encoder learn better, entities that exhibit long time presence, i.e., long temporal patterns that correspond to one output class like the sound of a car and the word ``car''. This need for DNNs has been identified at least since 2012~\cite{graves:2012:supervised} and the hierarchical sub-sampling networks were introduced, and also adopted more recently, e.g.~\cite{scheidegger:2017:eusipco}.

In this paper we employ the hierarchical sub-sampling networks and we present a novel approach for audio captioning methods employing multi-layered and RNN-based encoders. We draw inspiration from the empirical observation that each word in the caption corresponds to multiple time-steps of the input sequence to a DNN-based audio captioning method, and we hypothesize that the performance of the method could be enhanced by reducing the temporal length of the sequence after each RNN layer in an RNN-based encoder. To assess our hypothesis and our method, we employ a method that is not using temporal sub-sampling, and we alter it by solely employing the sub-sampling. The obtained results show that, indeed, temporal sub-sampling can enhance the performance of the audio captioning methods. 

The rest of the paper is organized as follows. In Section~\ref{sec:method} we present our method and the temporal sub-sampling method. In Section~\ref{sec:evaluation} we present the followed evaluation procedure, and the obtained results are in Section~\ref{sec:results}. Section~\ref{sec:conclusions} concludes the paper. 
\section{Proposed method}
\label{sec:method}
Our proposed method employs a multi-layered, bi-directional RNN-based encoder, an RNN-based decoder, accepts as an input a sequence of $T$ audio feature vectors with $F$ features, $\mathbf{X}\in\mathbb{R}^{T\times F}$, and outputs a sequence of $S$ vectors $\hat{\mathbf{Y}}=[\hat{\mathbf{y}}_{1},\ldots,\hat{\mathbf{y}}_{S}]$, where each vector $\hat{\mathbf{y}}\in[0,1]^{D}$ contains the predicted probability for each of the $D$ words available to the method, where $D$ is the amount of unique words in the captions of the training dataset. After each bi-directional RNN layer in the encoder (apart from the last one), our method applies a temporal sub-sampling of the output sequence of the bi-directional RNN layer, before use the sequence as an input for the next bi-directional RNN layer in the encoder. Figure~\ref{fig:method}  illustrates the method. Our method is otherwise the same as the baseline method of the DCASE 2020 automated audio captioning task (task 6)\footnote{\url{http://dcase.community/challenge2020/task-automatic-audio-captioning}}, enhanced with the of temporal sub-sampling of the latent representations in the encoder. 
\begin{figure}[!t]
    \centering
    \includegraphics[width=\columnwidth,trim={.1cm .2cm .1cm .2cm},clip]{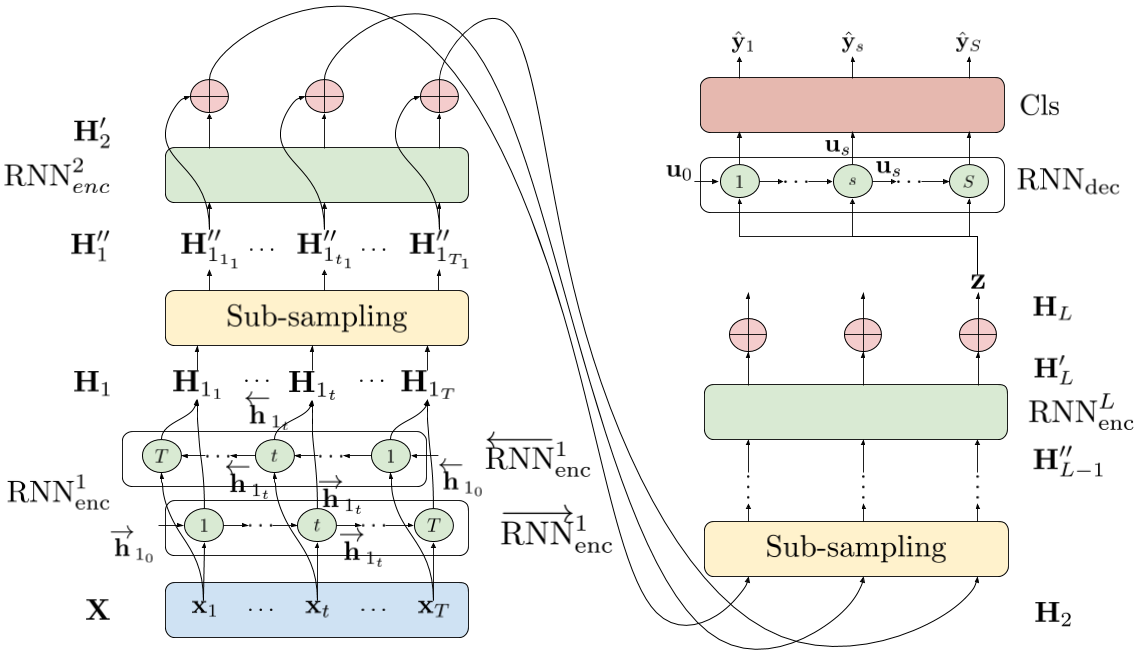}
    \caption{Illustration of the proposed method, with $L$ bi-directional RNN layers at the encoder. 
    Details are shown for $\text{RNN}^{l=1}_{\text{enc}}$ and for the result $l$ are similar. Details for sub-sampling are in Figure~\ref{fig:sub-sampling}.
    }
    \label{fig:method}
\end{figure}
\subsection{Encoder with temporal sub-sampling}\label{ssec:baseline}
Our encoder consists of $L_{\text{enc}}$ bi-directional RNNs, where $\overrightarrow{\text{RNN}}_{\text{enc}}^{l}$ and $\overleftarrow{\text{RNN}}_{\text{enc}}^{l}$ are the $l$-th forward and backward RNNs of the encoder, respectively. $\overrightarrow{\text{RNN}}_{\text{enc}}^{1}$ and $\overleftarrow{\text{RNN}}_{\text{enc}}^{1}$ process the input sequence $\mathbf{X}$ as
\begin{align}
    \overrightarrow{\mathbf{h}}_{1_{t}} =& \overrightarrow{\text{RNN}}_{\text{enc}}^{1}(\mathbf{x}_{t}, \overrightarrow{\mathbf{h}}_{1_{t-1}})\text{ and}\label{eq:forward-rnn}\\
    \overleftarrow{\mathbf{h}}_{1_{t}} =& \overleftarrow{\text{RNN}}_{\text{enc}}^{1}(\overleftarrow{\mathbf{x}}_{t}, \overleftarrow{\mathbf{h}}_{1_{t-1}})\label{eq:backward-rnn}\text{, }
\end{align}
\noindent
where $\overleftarrow{\mathbf{x}}_{t}$ is the $t$-th feature vector of the time-reversed $\mathbf{X}$, $\overrightarrow{\mathbf{h}}_{1_{t}},\overleftarrow{\mathbf{h}}_{1_{t}}\in[-1,1]^{\Xi}$ are the outputs of the $\overrightarrow{\text{RNN}}_{\text{enc}}^{1}$ and $\overleftarrow{\text{RNN}}_{\text{enc}}^{1}$, respectively, for the $t$-th time-step, $\overrightarrow{\mathbf{h}}_{1_{0}}=\overleftarrow{\mathbf{h}}_{1_{0}}=[0]^{\Xi}$, and $\Xi$ is the amount of output features of each of the RNNs of the $l$-th bi-directional RNN of the encoder. Then, the outputs of the $\overrightarrow{\text{RNN}}_{\text{enc}}^{1}$ and $\overleftarrow{\text{RNN}}_{\text{enc}}^{1}$, $\overrightarrow{\mathbf{h}}_{1_{t}}$ and $\overleftarrow{\mathbf{h}}_{1_{t}}$, respectively, are concatenated as
\begin{equation}\label{eq:bi-dir-concat}
    \mathbf{h}_{1_{t}} = [\overrightarrow{\mathbf{h}}_{1_{t}}^{\top}, \overleftarrow{\mathbf{h}}_{1_{t}}^{\top}]^{\top}\text{, }
\end{equation}
\noindent
and $\mathbf{H}_{1} = [\mathbf{h}_{1_{1}},\ldots,\mathbf{h}_{1_{T}}]$. Then, for $2\leq l < L$, our method applies a temporal sub-sampling to $\mathbf{h}_{l-1}$, as
\begin{equation}\label{eq:sub-sampling}
    \mathbf{H}''_{l-1} = \{\mathbf{h}^{l-1}_{iM+1}\}_{i=0}^{i=\lfloor (T_{l})/M\rfloor}\text{,}
\end{equation}
\noindent
where $T_{l}$ is the amount of time-steps of $\mathbf{H}''_{l-1}$, $M\in\mathbb{N}^{\star}$ is the sub-sampling factor, and $\lfloor\cdot\rfloor$ is the floor function. Figure~\ref{fig:sub-sampling}illustrates the sub-sampling process.

Then, $\mathbf{H}''_{l-1}$ is given as an input to $\overrightarrow{\text{RNN}}_{\text{enc}}^{l}$ and $\overleftarrow{\text{RNN}}_{\text{enc}}^{l}$, similarly to Eqs.~\eqref{eq:forward-rnn},~\eqref{eq:backward-rnn}, and~\eqref{eq:bi-dir-concat}, obtaining $\mathbf{H}'_{l}$. $\mathbf{H}_{l}$ is obtained by a residual connection between $\mathbf{H}''_{l-1}$ and $\mathbf{H}'_{l}$, as
\begin{equation}
    \mathbf{H}_{l} = \mathbf{H}'_{l} + \mathbf{H}''_{l-1}\text{,}
\end{equation}
\noindent
where $\mathbf{H}_{l}\in\mathbb{R}^{T_{l}\times\Delta}$ is the output of the $l$-th bi-directional RNN layer of the encoder. By utilizing Eq.~\eqref{eq:sub-sampling}, we enforce the RNNs of the encoder to squeeze the information in the input sequence to a smaller output sequence, effectively making the RNNs to learn a time-filtering and time-compression of the information in the input sequence~\cite{graves:2012:supervised,scheidegger:2017:eusipco}. This is can be proven beneficial in the case of audio captioning, since one output class (i.e. one word) corresponds to multiple input time-steps. By temporal sub-sampling, we are enforcing the encoder to express the learnt information with lower temporal resolution, effectively providing a shorter sequence but with each of its time-step to represent longer temporal patterns~\cite{graves:2012:supervised}. The above presented scheme of temporal sub-sampling, results in reducing the length of the input audio sequence to $M^{-L-1}$ times. For example, a sub-sampling factor of $M=2$ and $L=3$ RNN layers results in reducing the length of the input audio sequence $2^{2}$ times (a reduction of 75.0\%). 
\begin{figure}[!t]
    \centering
    \includegraphics[width=.6\columnwidth]{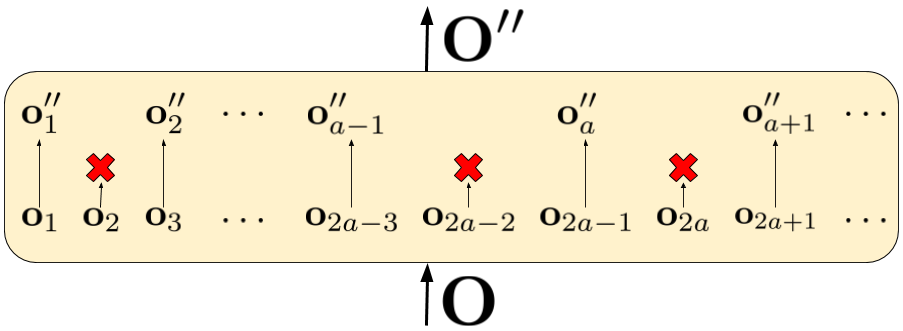}
    \caption{Illustration of the sub-sampling process with a factor $M=2$, with an input of a sequence with $A$ vectors of $B$ features, $\mathbf{O}\in\mathbb{R}^{A\times B}$, and an output of another sequence $\mathbf{O}''\in\mathbb{R}^{\lfloor A/M\rfloor\times B}$. With the red ``X'' we indicate the vectors that were discarded, according to the sub-sampling process.}
    \label{fig:sub-sampling}
\end{figure}

Finally, the output of the encoder. $\mathbf{z}\in\mathbb{R}^{\Delta}$, is formed as
\begin{equation}
    \mathbf{z} = \mathbf{H}_{L_{T_{L}}}\text{.}
\end{equation}
\noindent
That is, $\mathbf{z}$ is the last time-step of the output sequence $\mathbf{H}_{L}$ of the $L$-th bi-directional RNN of the encoder. 
\subsection{Decoder and optimization}
The decoder of our method consists of an RNN and a linear layer followed by a softmax non-linearity (the latter two will collectively referred to as classifier). The decoder takes as an input the $\mathbf{z}$ for every time step as
\begin{equation}\label{eq:dec-rnn}
    \mathbf{u}_{s} = \text{RNN}_{\text{dec}}(\mathbf{z}, \mathbf{u}_{s-1})\text{, }
\end{equation}
\noindent
where $\mathbf{u}_{s}\in[0, 1]^{\Psi}$ is the output of the $\text{RNN}_{\text{dec}}$ for the $s$-th time-step of the decoder, with $\mathbf{u}_{s}=[0]^{\Psi}$. $\mathbf{u}_{s}$ is used as an input to the classifier, $\text{Cls}$, as
\begin{equation}\label{eq:dec-cls}
    \hat{\mathbf{y}}_{s} = \text{Cls}(\mathbf{u}_{s})\text{.}
\end{equation}
The process described by Eqs~\eqref{eq:dec-rnn} and~\eqref{eq:dec-cls} is repeated until $s=S$ or $\mathbf{y}_{s}$ matches to a predefined symbol, depending if the decoder is in optimization or inference process, respectively. The encoder, the decoder, and the classifier are jointly optimized to minimize the binary cross-entropy at each time-step and between the predicted, $\hat{\mathbf{y}}_{s}$, and ground truth, $\mathbf{y}_{s}=[y_{s,1},\ldots,y_{s,D}]$, one-hot encoding of words. 
\section{Evaluation}\label{sec:evaluation}
To evaluate our method, we employ a freely and well-curated audio captioning dataset, called Clotho~\cite{drossos2020clotho}, and the baseline of the DCASE 2020 audio captioning task. For assessing the performance of our method, we use machine translation and captioning metrics, employed by most of the existing audio captioning work.
\subsection{Dataset and data pre-processing}
\label{ssec:dataset}
Clotho contains audio clips of CD quality (44.1 kHz sampling rate, 16-bit sample width), and 5 captions for each audio clip. The time duration of the audio clips ranges from 15 to 30 seconds, and the amount of words in each caption ranges from eight to 20 words. Clotho provides three splits for developing audio captioning methods, namely development, evaluation, and testing. The development and evaluation splits are freely available online\footnote{\url{https://zenodo.org/record/3490684}}, while the testing split is withheld for scientific challenges. In this work, we employ the development and evaluation splits of Clotho, having 2893 and 1045 audio clips, yielding 14465 and 5225 captions, respectively. We choose Clotho because it is built to offer audio content diversity, and extra care has been taken for eliminating spelling errors, named entities, and speech transcription in the captions. Additionally, Clotho is already employed at the DCASE 2020 audio captioning task$^{1}$. 

We extract $F=64$ log-scaled mel-band energies from each of the 4981 audio clips of Clotho, using an 1024-sample long window (approximately 23 ms) with 50\% overlap, and the Hamming windowing function. This results in having sequences from $T=1292$ to $T=2584$ audio feature vectors, for audio clips of 15 and 30 seconds duration, respectively. We process the captions of Clotho, starting by appending $\left<\text{eos}\right>$ to all captions, where $\left<\text{eos}\right>$ is a special token that signifies the end of the sequence (i.e. the caption). Then, we identify the set of words in the captions, include $\left<\text{eos}\right>$ in that set as well, and represent each element of that set (called a token from now on) with an one-hot encoding vector $\mathbf{y}=\{0,1\}^{D}$. This process yields a sequence of $S=8$ to $S=21$ one-hot encoded tokens for each caption. To implement the above, we employed the freely available code from the DCASE 2020 audio captioning task\footnote{\url{https://github.com/audio-captioning/clotho-dataset}}. We use each audio clip with all of its corresponding captions as separate input-output examples, resulting to a total of 14465 and 5225 input-output examples for the development and evaluation splits, respectively. We use each of the sequences of audio feature vectors in the splits as our $\mathbf{X}$ and each of the corresponding sequences of one-hot encoded tokens as our $\mathbf{Y}$.
\subsection{Hyper-parameters and training procedure}\label{ssec:training}
To optimize our method and fine tune its hyper-parameters, we employ the development split. By empirical observation on the values of the loss for the development split, we decided to round the loss value to three decimal digits and stop the training if the loss did not improve for 100 consecutive epochs. For the training of our method, we employed a batch size of 16 (mainly due to computational resources constraints). To apply a uniform $T$ and $S$ in a batch, we calculate the maximum $T$ and $S$ in the batch and we pre-pend vectors of $\{0\}^{F}$ to each $\mathbf{X}$ and append the one-hot encoded vector of $\left<\text{eos}\right>$ to every $\mathbf{Y}$, in order to make them have the $T$ and $S$ equal to the maximum $T$ and $S$ in the same batch. Additionally, we observed that there is a considerable imbalance at the frequency of appearance of the tokens at the captions. That is, some $w_{i}$, for example ``a''/``an'' and ``the'', appear quite frequently (e.g. over 4000 times) at the captions, but some appear quite fewer times, e.g. five. To overcome this imbalance, each token, $w_{i}$, is inversely weighted by its frequency in the dataset, resulting in the following loss formulation
\begin{equation}
    \mathcal{L}'(\hat{\mathbf{y}}_{s}, \mathbf{y}_{s}) = \Phi_{s}\mathcal{L}(\hat{\mathbf{y}}_{s}, \mathbf{y}_{s})\text{,}
\end{equation}
\noindent
where $\Phi_{s}$ is a weight for the loss calculation of the token represented by $\mathbf{y}_{s}$. But due to the quite large frequency of tokens like ``a'', we observed that $\Phi_{s}$ could get values as low as $1e-5$, which when compared to a value of $\Phi_{s}=1$ for the non-frequent tokens, has a great difference. This difference at the values of $\Phi_{s}$ results in hampering significantly the learning of the frequent tokens and, in addition, will not ever contribute to the training of our method since we round $\mathcal{L}'$ to three decimal digits. Thus, we employed a clamping of $\Phi_{s}$ as
\begin{equation}
    \Phi_{s} = \begin{cases}
    \frac{\min(f_{w})}{f_{w_{s}}}& \text{if }~\frac{\min(f_{w})}{f_{w_{s}}}\geq\beta\text{,}\\
    \beta& \text{otherwise}
    \end{cases}\text{,}
\end{equation}
\noindent
where $\beta$ is a hyper-parameter that we set to $5e-1$ by following the above described process, $f_{w_{s}}$ is the frequency of the $w_{s}$ token in the development split, $w_{s}$ is the token that the $\mathbf{y}_{s}$ corresponds to, and $\min(f_{w})$ is the minimum frequency of all tokens in the Clotho dataset. We follow the baseline method of the DCASE 2020 audio captioning task, and we use $L=3$, with $\Xi=256$ and $\Psi=256$, which are the same as in the baseline of DCASE 2020 audio captioning task. According to Clotho, $D=4366$. We optimize the parameters of our method using $\mathcal{L}'$ and the Adam optimizer~\cite{kingma2014adam}, with a learning rate of $1e-4$ and the values for $\beta_{1}$ and $\beta_{2}$ that are reported to the corresponding paper~\cite{kingma2014adam}, and we employ dropout with a probability of $p=0.25$ between $\text{RNN}^{1}_{\text{enc}}$ and $\text{RNN}^{2}_{\text{enc}}$, and between $\text{RNN}^{2}_{\text{enc}}$ and $\text{RNN}^{3}_{\text{enc}}$.

We choose hyper-parameters that yielded the lowest loss value for the development split, following the above mentioned policy of stopping the training process and implementing a random search over the combinations of $\Phi_{s}$ and learning rate of Adam. To evaluate the impact of the sub-sampling, we employ four different values for $M$, namely 2, 4, 8, and 16. Finally, the total amount of parameters of our method is 4 573 711, and the code of our method is based on the PyTorch framework and is freely available online\footnote{\url{https://github.com/DK-Nguyen/audio-captioning-sub-sampling}}.
\subsection{Evaluation and metrics}\label{ssec:metrics}
We assess the performance of our method by using the above mentioned processes and hyper-parameters, and employing the metrics used in the DCASE 2020 audio captioning task.  Each metric is calculated using the the predicted sequence of words $\hat{\mathbf{Y}}$ for a $\mathbf{X}$, and all the ground truth sequences $\mathbf{Y}$ for the same $\mathbf{X}$. We compare the obtained values against the ones reported by the baseline method of the DCASE 2020 audio captioning task, which is our method with $M=1$. The calculation of the metrics is performed using the available tools for DCASE 2020 audio captioning task\footnote{\url{https://github.com/audio-captioning/caption-evaluation-tools}}.

Specifically, we use the machine translation metrics $\text{BLEU}_{n}$, $\text{ROUGE}_{\text{L}}$, and METEOR, and the captioning metrics CIDEr, SPICE, and SPIDEr. $\text{BLEU}_{n}$ is a precision-based metrics that measures a weighted geometric mean of modified precision of $n$-grams between predicted and ground truth captions~\cite{papineni2002bleu}. $\text{ROUGE}_{\text{L}}$~\cite{lin2004rouge} calculates an F-measure using a longest common sub-sequence (LCS) between predicted and ground truth captions, and METEOR~\cite{lavie2007meteor} measures a harmonic mean of the precision and recall for segments between predicted and ground truth captions, which is shown to have high correlation with quality human-level translation. CIDEr~\cite{vedantam2015cider} calculates a weighted cosine similarity using term-frequency inverse-document-frequency (TF-IDF) weighting for $n$-grams, and SPICE~\cite{anderson2016spice} measures the ability of the predicted captions to recover from the ground truth captions, objects, attributes, and the relationship between them. Finally, SPIDEr is a weighted mean between CIDEr and SPICE, exploiting the advantages of both metrics~\cite{2017:liu:iccv}. We assess the performance of our method versus the DCASE 2020 audio captioning task method using the values of the above mentioned metrics, evaluated on the evaluation split of Clotho. 
\begin{table}[!t]
\centering
\caption{Results for the baseline method, i.e. $M=1$, and our proposed method with sub-sampling factor $M=\{2, 4, 8, 16\}$.}
\label{tab:results}
\resizebox{\columnwidth}{!}{%
\begin{tabular}{lccccc}
\textbf{Metric} & 
$\boldsymbol M \boldsymbol=\boldsymbol1$ & 
$\boldsymbol M \boldsymbol=\boldsymbol2$ &
$\boldsymbol M \boldsymbol=\boldsymbol4$ & 
$\boldsymbol M \boldsymbol=\boldsymbol8$ &
$\boldsymbol M \boldsymbol=\boldsymbol16$ \\
\hline
$\text{BLEU}_{1}$ & 0.389 & \textbf{0.426} & 0.418 & 0.417 & \textbf{0.426} \\ 
$\text{BLEU}_{2}$ & 0.136 & 0.151 & 0.151 & \textbf{0.154} & 0.147 \\
$\text{BLEU}_{3}$ & 0.055 & 0.058 & 0.061 & \textbf{0.063} & 0.058 \\
$\text{BLEU}_{4}$ & 0.015 & 0.020 & 0.018 & \textbf{0.025} & 0.022 \\
$\text{ROUGE}_{\text{L}}$ & \textbf{0.262} & 0.274 & 0.275 & 0.274 & 0.274 \\
METEOR & 0.084 & \textbf{0.092} & 0.091 & 0.089 & 0.090 \\
CIDEr & 0.074 & 0.092 & 0.090 & \textbf{0.093} & \textbf{0.093} \\
SPICE & 0.033 & \textbf{0.040} & 0.037 & \textbf{0.040} & 0.036 \\
\hline
SPIDEr & 0.054  & 0.066 & 0.064 & \textbf{0.067} & 0.064
\end{tabular}
}
\end{table}
\section{Results and discussion}\label{sec:results}
In Table~\ref{tab:results} are the values of the employed metrics for the evaluation split of Clotho. As can be seen from Table~\ref{tab:results}, using a sub-sampling factor $M\geq2$ always improves the values of the metrics. This fact clearly indicates that our proposed method of sub-sampling benefits the performance of audio captioning methods. The maximum value of SPIDEr is obtained for $M=8$, is 0.067, and can be mainly attributed to the better SPICE score than the value of $M=8$ yields. Though, the values of the metrics for the different sub-sampling factors, do not exhibit some systematic behaviour. That is, increasing above 2, i.e. $M>2$ does not result in increasing or decreasing the performance. This could be attributed to the fact that the employed method employs a fixed-length output from the encoder. Thus the increased impact from reducing more the length of the sequence, cannot be observed since the output of the encoder is always a fixed-length vector. This fact strongly indicates that the impact of the increased $M$ might be more visible in an audio captioning method, that employs an alignment mechanism which uses the whole output sequence of the encoder, and not only a fixed-length vector (e.g. attention).
\begin{table}[!t]
    \centering
    \caption{Reduction of the sequence length (in percentages) compared to the input audio, required time for predictions on Clotho evaluation split, and minimum and maximum resulting amount of time-steps ($T^{\text{min}}_{L}$ and $T^{\text{max}}_{L}$, respectively), according to sub-sampling factor $M=\{2, 4, 8, 16\}$, and $L=3$.}
    \label{tab:length-reduction}
    \begin{tabular}{S[table-format=2.0]S[table-format=4.0]S[table-format=4.0]S[table-format=2.2]S[table-format=2.2]}
         $\boldsymbol M$ & $\boldsymbol{T^{\text{\textbf{min}}}_{L}}$ & $\boldsymbol{T^{\text{\textbf{max}}}_{L}}$ & \textbf{Reduction in length} & \textbf{Time (sec)}\\
         \hline
          1  & 1292 & 2584 & 00.00\% & 58.81\\
          2  &  323 &  646 & 75.00\% & 34.13\\
          4  &   80 &  161 & 93.75\% & 25.67\\
          8  &   20 &   40 & 98.43\% & 21.73\\
         16  &    5 &   10 & 99.60\% & 20.42
    \end{tabular}
\end{table}

In Table~\ref{tab:length-reduction} is the resulting reduction in the time needed for obtaining all the predicted outputs using the Clotho evaluation split, and the resulting length of the output sequence of the encoder compared to the input sequence $\mathbf{X}$. As can be observed from Table~\ref{tab:length-reduction}, the increase in $M$ has also a clear impact at the time needed for obtaining the predicted captions. This is to be expected, since with temporal sub-sampling, the output of the encoder is 99.6\% shorter compared to the length of the input audio. 

Finally, an example of the output of our method with $M=8$ is ``\emph{a person is walking a through something and}'', for the file of the Clotho evaluation split ``clotho\_file\_01 A pug struggles to breathe 1\_14\_2008'' and with ground truth captions like ``a man walking who is blowing his nose hard and about to sneeze'' and ``a small dog with a flat face snoring and groaning''. Another example is the predicted caption ``\emph{a group of of birds and birds a}'' for the file ``clotho\_file\_sparrows'' and with ground truth captions like ``a flock of birds comes together with a lot of chirping'' and ``birds sing in different tones while in a large group''. As it can be seen, out method manages to identify the sources and the actions, but it lacks in the language modelling (LM). The latter fact is to be expected, since our method did not focused on the LM perspective, e.g. by using attention or explicit LM. 
\section{Conclusions and future work}\label{sec:conclusions}
In this paper we presented an approach for audio captioning that utilizes temporal sub-sampling, given the empirical observation that a word in a general audio caption refers to a sequence of audio samples. Our approach is focusing on methods that use a multi-layered and RNN-based encoder, utilizing a temporal sub-sampling of the output sequence of each RNN layer of the encoder. 
We evaluated our approach using the freely available audio captioning dataset, Clotho, using multiple factors of sub-sampling. The obtained results clearly indicate that temporal sub-sampling can benefit the audio captioning methods. We observed an increase at all metrics and with all sub-sampling factors greater than 2, compared to the case of not using sub-sampling (i.e. $M=1$). The maximum benefit was observed for $M=8$, where an increase of 1.3 is observed for SPIDEr. From the variation of the values of the utilized metrics, according to the different sub-sampling factors, we hypothesize that the temporal sub-sampling might have a more pronounced effect when is employed in a method that uses an alignment mechanism like attention. Though, more research is needed for verifying or disproving that. 
\bibliographystyle{IEEEtran}
\bibliography{refs}

\begin{thebibliography}{10}
\providecommand{\url}[1]{#1}
\def\UrlFont{\rmfamily}
\providecommand{\newblock}{\relax}
\providecommand{\bibinfo}[2]{#2}
\providecommand\BIBentrySTDinterwordspacing{\spaceskip=0pt\relax}
\providecommand\BIBentryALTinterwordstretchfactor{4}
\providecommand\BIBentryALTinterwordspacing{\spaceskip=\fontdimen2\font plus
\BIBentryALTinterwordstretchfactor\fontdimen3\font minus
  \fontdimen4\font\relax}
\providecommand\BIBforeignlanguage[2]{{%
\expandafter\ifx\csname l@#1\endcsname\relax
\typeout{** WARNING: IEEEtran.bst: No hyphenation pattern has been}%
\typeout{** loaded for the language `#1'. Using the pattern for}%
\typeout{** the default language instead.}%
\else
\language=\csname l@#1\endcsname
\fi
#2}}

\bibitem{drossos2017automated}
K.~Drossos, S.~Adavanne, and T.~Virtanen, ``Automated audio captioning with
  recurrent neural networks,'' in \emph{2017 IEEE Workshop on Applications of
  Signal Processing to Audio and Acoustics (WASPAA)}, 2017, pp. 374--378.

\bibitem{lipping:2019:dcase}
S.~Lipping, K.~Drossos, and T.~Virtanen, ``Crowdsourcing a dataset of audio
  captions,'' in \emph{Detection and Classification of Acoustic Scenes and
  Events ({DCASE}) 2019}, Oct. 2019.

\bibitem{wu:2019:icassp}
M.~{Wu}, H.~{Dinkel}, and K.~{Yu}, ``Audio caption: Listen and tell,'' in
  \emph{ICASSP 2019 - 2019 IEEE International Conference on Acoustics, Speech
  and Signal Processing (ICASSP)}, 2019, pp. 830--834.

\bibitem{drossos2020clotho}
K.~Drossos, S.~Lipping, and T.~Virtanen, ``Clotho: An audio captioning
  dataset,'' in \emph{ICASSP 2020-2020 IEEE International Conference on
  Acoustics, Speech and Signal Processing (ICASSP)}.\hskip 1em plus 0.5em minus
  0.4em\relax IEEE, 2020, pp. 736--740.

\bibitem{kim:2019:acl}
C.~D. Kim, B.~Kim, H.~Lee, and G.~Kim, ``{A}udio{C}aps: Generating captions for
  audios in the wild,'' in \emph{Proceedings of the 2019 Conference of the
  North {A}merican Chapter of the Association for Computational Linguistics:
  Human Language Technologies, Volume 1 (Long and Short Papers)}, Jun. 2019,
  pp. 119--132.

\bibitem{cho2014learning}
\BIBentryALTinterwordspacing
K.~Cho, B.~van Merri{\"e}nboer, C.~Gulcehre, D.~Bahdanau, F.~Bougares,
  H.~Schwenk, and Y.~Bengio, ``Learning phrase representations using {RNN}
  encoder{--}decoder for statistical machine translation,'' in
  \emph{Proceedings of the 2014 Conference on Empirical Methods in Natural
  Language Processing ({EMNLP})}.\hskip 1em plus 0.5em minus 0.4em\relax Doha,
  Qatar: Association for Computational Linguistics, Oct. 2014, pp. 1724--1734.
  [Online]. Available: \url{https://www.aclweb.org/anthology/D14-1179}
\BIBentrySTDinterwordspacing

\bibitem{bahdanau2014neural}
D.~Bahdanau, K.~Cho, and Y.~Bengio, ``Neural machine translation by jointly
  learning to align and translate,'' in \emph{Proceedings of the International
  Conference on Learning Representation ({ICLR})}, 2014.

\bibitem{seo:2016:iclr}
M.~J. Seo, A.~Kembhavi, A.~Farhadi, and H.~Hajishirzi, ``Bidirectional
  attention flow for machine comprehension,'' in \emph{Proceedings of the
  International Conference on Learning Representation ({ICLR})}, vol.
  abs/1611.01603, 2016.

\bibitem{vaswani:2017:neurips}
A.~Vaswani, N.~Shazeer, N.~Parmar, J.~Uszkoreit, L.~Jones, A.~N. Gomez,
  {\L}.~Kaiser, and I.~Polosukhin, ``Attention is all you need,'' in
  \emph{Proceedings of the 31st International Conference on Neural Information
  Processing Systems}.\hskip 1em plus 0.5em minus 0.4em\relax Red Hook, NY,
  USA: Curran Associates Inc., 2017, p. 6000–6010.

\bibitem{graves:2012:supervised}
\BIBentryALTinterwordspacing
A.~Graves, \emph{Supervised Sequence Labelling with Recurrent Neural Networks},
  ser. Studies in Computational Intelligence.\hskip 1em plus 0.5em minus
  0.4em\relax Springer Berlin Heidelberg, 2012. [Online]. Available:
  \url{https://books.google.fi/books?id=wpb-CAAAQBAJ}
\BIBentrySTDinterwordspacing

\bibitem{scheidegger:2017:eusipco}
F.~{Scheidegger}, L.~{Cavigelli}, M.~{Schaffner}, A.~C.~I. {Malossi},
  C.~{Bekas}, and L.~{Benini}, ``Impact of temporal subsampling on accuracy and
  performance in practical video classification,'' in \emph{2017 25th European
  Signal Processing Conference (EUSIPCO)}, 2017, pp. 996--1000.

\bibitem{kingma2014adam}
D.~P. Kingma and J.~Ba, ``Adam: A method for stochastic optimization,'' in
  \emph{Proceedings of the International Conference on Learning Representation
  ({ICLR})}, 2014.

\bibitem{papineni2002bleu}
K.~Papineni, S.~Roukos, T.~Ward, and W.-J. Zhu, ``Bleu: a method for automatic
  evaluation of machine translation,'' in \emph{Proceedings of the 40th annual
  meeting on association for computational linguistics}.\hskip 1em plus 0.5em
  minus 0.4em\relax Association for Computational Linguistics, 2002, pp.
  311--318.

\bibitem{lin2004rouge}
\BIBentryALTinterwordspacing
C.-Y. Lin, ``{ROUGE}: A package for automatic evaluation of summaries,'' in
  \emph{Text Summarization Branches Out}.\hskip 1em plus 0.5em minus
  0.4em\relax Barcelona, Spain: Association for Computational Linguistics, July
  2004, pp. 74--81. [Online]. Available:
  \url{https://www.aclweb.org/anthology/W04-1013}
\BIBentrySTDinterwordspacing

\bibitem{lavie2007meteor}
A.~Lavie and A.~Agarwal, ``Meteor: An automatic metric for mt evaluation with
  high levels of correlation with human judgments,'' in \emph{Proceedings of
  the second workshop on statistical machine translation}, 2007, pp. 228--231.

\bibitem{vedantam2015cider}
R.~Vedantam, C.~Lawrence~Zitnick, and D.~Parikh, ``{CIDEr}: Consensus-based
  image description evaluation,'' in \emph{Proceedings of the IEEE conference
  on computer vision and pattern recognition ({CVPR})}, 2015, pp. 4566--4575.

\bibitem{anderson2016spice}
P.~Anderson, B.~Fernando, M.~Johnson, and S.~Gould, ``Spice: Semantic
  propositional image caption evaluation,'' in \emph{European Conference on
  Computer Vision}.\hskip 1em plus 0.5em minus 0.4em\relax Springer, 2016, pp.
  382--398.

\bibitem{2017:liu:iccv}
S.~{Liu}, Z.~{Zhu}, N.~{Ye}, S.~{Guadarrama}, and K.~{Murphy}, ``Improved image
  captioning via policy gradient optimization of spider,'' in \emph{2017 IEEE
  International Conference on Computer Vision (ICCV)}, 2017, pp. 873--881.

\end{thebibliography}
\end{sloppy}
\end{document}